\documentclass[reprint, amssymb, amsmath, aip,cha,]{revtex4-1}

\usepackage{docs}%
\usepackage{bm}%
\usepackage{graphicx}
\usepackage{dcolumn}
\usepackage{bm}

\usepackage{wasysym}
\usepackage{textcomp}
\usepackage{epstopdf}
\usepackage{stmaryrd}
\usepackage[colorlinks=true,linkcolor=blue]{hyperref}%
\expandafter\ifx\csname package@font\endcsname\relax\else
 \expandafter\expandafter
 \expandafter\usepackage
 \expandafter\expandafter
 \expandafter{\csname package@font\endcsname}%
\fi
\begin{document}

\title{Preferentially orientated E-beam TiN thin films using focused jet of nitrogen gas}

\author{R. Ramaseshan}
\email{seshan@igcar.gov.in}
\affiliation{Thin Film and Coatings Section, Surface and Nanoscience Division, Materials Science Group, Indira Gandhi Centre for Atomic Research, Kalpakkam 603102, India.}

\author{Feby Jose, S. Rajagopalan and S. Dash}

\date{\today}

\begin{abstract}
A modified electron beam evaporator has been used judiciously to synthesize TiN thin films with (111) preferred orientation. This new design involved in creating local plasma by accelerating the secondary electrons emitted from the evaporating ingot by a positively biased semi-cylindrical anode plate kept in the vicinity and a jet of $N_{2}$ gas has been focused towards the substrate as a reactive gas. We have observed a preferred orientation (111) with $\ 25^{\circ}$ angle to the surface normal and this was confirmed by pole figure analysis. The phenomenon of preferred orientation (111) has been explained based on the rate of evaporation. The residual stress by the classical sin$^2\psi$ technique did not yield any tangible result due to the preferred orientation. The hardness and modulus measured by nanoindentation technique was around 19.5~GPa and 214~GPa. The continuous multicycle indentation test on these films exhibited a stress relaxation. 

\end{abstract}

\maketitle


\section{Introduction}

Polycrystalline TiN films, belongs to the family of refractory transition metal nitrides exhibit characteristics of both covalent compounds and metals, such as high melting point, thermodynamic stability, high hardness, thermal conductivity which are necessary for the mechanical components such as cutting tools, forming tools, etc.\cite{Dong, Wagner, Ma, Randall, Lackner} In addition, TiN thin films have been used for cosmetic ersatz gold surfaces such as watch bezels, watch bands, wavelength selective transparent optical films, diffusion barrier in integrated circuits, and as energy saving coatings for windows due to its strong infrared reflection.\cite{Allan,Yustea,Elias} Preferred oriented thin films largely affect the properties, performance and reliability of the components than the polycrystalline thin films.\cite{Deniz, Abadias2004} 

There are several techniques reported for synthesizing TiN in polycrystalline as well as highly oriented films includes reactive evaporation, \cite{Hakansson, Mori} magnetron sputtering, \cite{Ma, Chen} pulsed laser deposition (PLD) \cite{Lackner} and cathodic arc evaporation, \cite{Krella} etc. Diatomic fcc structure TiN has a favored (200) orientation which is substantially supported  thermodynamically due to their low surface energy.\cite{MahieuTSF} The ion bombardment of the growing film in the vacuum based plasma assisted process, the presence of high-density plasma is identified as one of the most relevant deposition parameters and its variation may drastically alter the film structure and properties.\cite{Ehiasarian, seshan, Abadias} Preferred crystalline growth and residual stress represent important structural features of polycrystalline thin films that influence the functionality significantly.\cite{Mahieu} An important question about the interrelation between preferred orientation and residual stress that changes the properties of these thin films has been explained by Shin \emph{et al}. \cite{Shin}
	
	The purpose of this work is first to examine the effect of the positive bias applied to enhance the formation of plasma near the vicinity and the gas feed direction on the growth of TiN films with preferential orientation. 

\section{Experimental}

\begin{table}[b]
\begin{center}
 \begin{tabular}{|l|c|}
  \hline  
  Electron Beam Power (W) &	250  \\\hline
  Working Pressure (mbar)&	5 x$ 10^{-4}$ \\\hline
  Cylindrical plate potential (V) &	+100 DC \\\hline
  Substrate biasing voltage (V) &	-500 DC \\\hline
  Source-substrate distance (mm) &	100 \\\hline
  Plasma current (A) &	5 \\\hline
  Deposition Temperature (K) &	525  \\
  \hline
 \end{tabular}\
 \caption{Deposition Parameters}
 \label{table:Depo.para}
\end{center}
 \end{table}
An electron beam evaporator (in house assembled) was improvised for synthesizing ARE based nitrides. A schematic of the deposition process enhanced by the discharge assembly setup is depicted in Fig.1. The vacuum chamber was pumped down to a base pressure of 1 x $ 10^{-6}$~mbar by turbo molecular pump. A Temescal SIMBA 2 power supply with electron beam rastering capability was used to evaporate titanium (4N pure) ingot. Prior to evaporation, the ingots were surface melted in-situ and degassed. During deposition, a constant flow of 5~N pure nitrogen was allowed (100 SCCM) and it was directed towards the substrate in the vicinity of the discharge assembly with an angle $\ 25^{\circ}$ to the substrate surface normal. The typical deposition parameters are listed in Table.\ref{table:Depo.para}.

\begin{figure}[ht]
\includegraphics[width=0.42\textwidth]{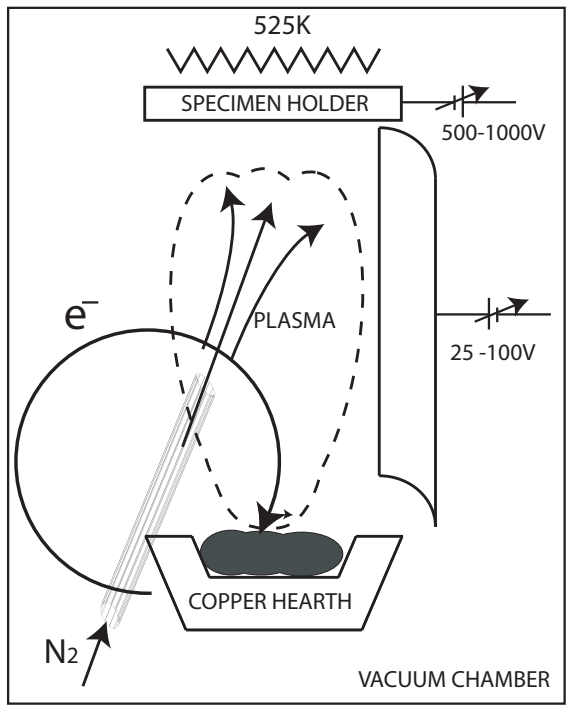}
\caption{\label{Fig1}Schematic of ARE-EB-PVD set up}
\end{figure}

A stainless steel semi cylindrical plate having 25~mm radius and 50~mm length was placed 20~mm above the edge of the molten source. A positive DC bias of 100~V was applied to the cylindrical plate. The secondary electrons emanating from the ingot by electron beam impact is accelerated towards the semi cylindrical anode. These secondary electrons induce ionization of Ti vapor and nitrogen molecules. Axi-symmetric plasma engulfs the space between the molten source and the substrate region, which is sustained by the high constant current supply of the anode bias. Throughout the deposition a constant plasma current of 5~A was maintained. The ionized species are transported towards a proximally held substrate kept at a negative DC bias of 500~V, (low current). The advantage of this geometry for plasma creation is that, it is confined plasma, which enhances the purity of the films. \cite{Wu} The deposition of crystalline TiN films was carried out on micro slide (glass slides) substrate by maintaining the substrate temperature of 525~K. The deposition rate of 20~nm/min was achieved with the above mentioned conditions. 

Identification of crystalline phases and crystal structure studies were carried out by a STOE diffractometer in the GIXRD mode at glancing angle of $1^{\circ}$. The pole figure and residual stress mapping of the thin films were analyzed by X-ray diffractometer (D8 Advance, M/s. Bruker, Germany) at an incidence angle of $0.5^{\circ}$. Compositional homogeneity with respect to depth upto substrate interface was studied by Secondary Ion Mass Spectrometry (SIMS) (IMS 4f, M/s. Cameca, France) instrument. The primary ion (Cs$^+$) beam with impact energy of 1.75~keV was used for sputtering with a beam current of 10 nA. The primary ion beam was rastered over an area of 200 x 200~$\mu$m and the secondary species were collected from a central region of 60~$\mu$m diameter. The SIMS crater depth and thickness of the films were measured with a surface profiler (DEKTAK 6M, M/s. Veeco, USA).

Surface morphology of TiN films was studied by using SIS AFM in contact mode. The hardness studies were carried out using nano-indentation system (Open Platform, M/s. CSM, Switzerland) using a Berkovich indenter, according to the ISO 14577-1 standard with a loading duration to peak load in 30 s and unload at the same rate as loading. The hardness and Young’s modulus were calculated from the load-displacement curve with the help of Oliver and Pharr formalism\cite{Oliver}. We have used continuous multi-cycle mode (CMC) technique with a load range of 1 - 20~mN with a progressive load increment of 1~mN at the same place (20~cycles) to observe any change in the hardness. The unloading was allowed up to 10\% of the corresponding applied load to maintain the contact between tip and surface.
 
\section{Results and Discussion}

\begin{table*}[ft]
\begin{center}
 
 \begin{tabular}{|c|c|c|c||c|c|c|}
  \hline
  \hline
  
Miller Indices & Intensity (JCPDS)\cite{JCPDS} & Normalized Intensity & d-reference (nm)& d-measured (nm) & $\Delta$d\% & $\gamma_{\mathrm{hkl}}^{\mathrm{*}}$ \\\hline

(111) & 72 & 84.3 &	0.24491 & 0.24214	 & 1.13 & 30 \\  \hline
(200) & 100 & 100.0 &	0.21207 & 0.20865	 & 1.61 & 36 \\  \hline 
(220) & 45 & 56.6 &	0.14996 & 0.14750	 & 1.64 & 20 \\  \hline 
(311) & 19 & 19.6 &	0.12789 & 0.12642	 & 1.15 & 7 \\  \hline 
(222) & 12 & 15.0 &	0.12244 & 0.12107	 & 1.12 & 5 \\  \hline

 \end{tabular}\
 \caption{Relative intensity, d-spacing and  texture coefficient of observed peaks}
 \label{table:xray}
\end{center}
 \end{table*}

In the present deposition set up, enhanced reactivity and high adatom energy imparted by localized plasma and substrate bias ensures a complete reaction between Ti and N ions and the formation of TiN films. By varying the constant current supply to the anode plate and evaporation rate, different deposition rates can be realized. However, when we increased the deposition rate by increasing the titanium evaporation rate alone by adjusting the electron beam power, both Ti and TiN were formed together. So, we have synthesized TiN thin films using the parameters mentioned in the experimental procedure. The GIXRD pattern of TiN films deposited on the micro slides by ARE based EB-PVD is shown in Fig.2. This diffraction pattern matches exactly with the reported JCPDS-ICDD data file, \cite{JCPDS} albeit systematic shift of all Bragg reflections to higher angle. This systematic shift in angle corresponds to the lattice contraction that is attributed to compressive residual stress in the film. The lattice constant of this film calculated from the highest peak of the x-ray diffraction is 0.41955~nm, whereas the reference value is 0.42410~nm. The high $\Delta$d\% (1.64) indicating there is a large compressive stress in the film compared to standard TiN. \cite{Hakansson} Such residual strain are expected in reactively deposited films due to (i) the thermal mismatch between thin film phases and the substrate, (ii) the lattice mismatch between the two, (iii) point defects due to non – stoichiometry, (iv) preferred orientations if any and (v) presence of dislocations. However, the presence of compressive residual stress in the film enhances its life by hardening the surface and controlling the crack growth.

\begin{figure}[h]
\includegraphics[width=0.42\textwidth]{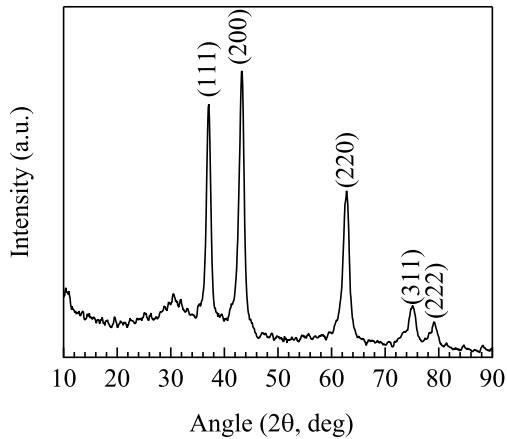}
\caption{\label{Fig2}GIXRD profile of ARE-EB-PVD-TiN film}
\end{figure}

\begin{figure}[h]
\includegraphics[width=0.42\textwidth]{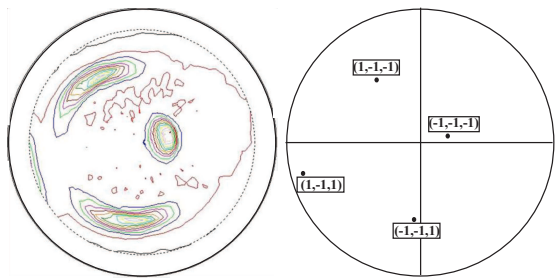}
\caption{\label{Fig3}(Color online) Pole figure of (111) plane with $25^{\circ}$ off in the $\psi$ with respect to surface normal and (111) plane generated by CaRine}
\end{figure}

\begin{figure}[h]
\includegraphics[width=0.42\textwidth]{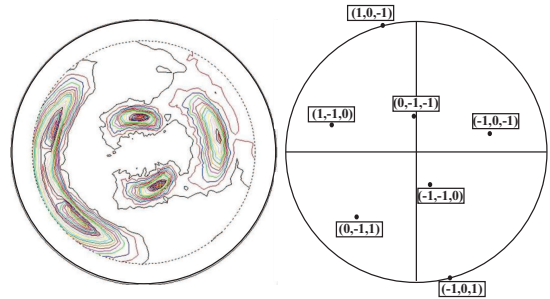}
\caption{\label{Fig4}(Color online) Pole figure of (220) plane and (220) plane generated by CaRine}
\end{figure}

 Although the relative diffracted intensities are not quantitative in GIXRD mode, these films can be inferred to have ideally random polycrystalline nature as learned from the relative intensities comparison given in the Table.\ref{table:xray} with slight prominence to (111) and (200) orientations. The texture coefficient of observed peaks are calculated using the $\gamma_{\mathrm{hkl}}^{\mathrm{*}}$ = $\textit{I}_{\mathrm{hkl}}/\sum\textit{I}_{\mathrm{hkl}}$, where $\textit{I}_{\mathrm{hkl}}$ is the intensity of the specific peak and $\sum\textit{I}_{\mathrm{hkl}}$ is sum of intensities of all observed TiN peaks.  The atomic packing density of (200) plane (4~atoms/$a^2$) is higher than (111) plane (2.3~atoms/$a^2$) and are higher than all other miller index planes. In general, the (200) planes of rock salt type structure have the lowest surface energy whereas (111) planes have the lowest strain energy. \cite{Xu, Dongdong} When the rate of deposition is low, the adatoms can get enough time to diffuse and react. In this process, they release self-reaction energy and reach the lowest surface energy. In reactive electron beam evaporation Ti ions are preferentially excited \cite{Xiao} and the substrate bias leading to self-ion bombardment further assists atomic rearrangement. The films with slight preference to orientation of high atomic density planes will pose higher resistance against oxidation. The texture analysis by LEPTOS software has shown a clear preferential orientation of (111) and (220) planes. Fig.3 in particular shows a huge texture with (111) planes inclined about $\ 25^{\circ}$ from the surface normal and its corresponding simulated pattern of (111). A similar preferred orientation was observed by some researchers, where the Ion beam assisted deposition and magnetron sputtering has been used to get the same with an angle. \cite{Jia, Birkholz} 
 
Similarly, fig.4 shows the (220) oriented planes also grown at the expense of (200) planes in this system along with its simulated pattern of (220). In this deposition system a tube for N$_2$ gas focused at an angle $\ 25^{\circ}$ to the surface normal. Such configuration enhanced the formation of (111) planes as well as the tilt of this plane to surface normal along with the low rate of deposition. 

The classical sin$^2\psi$ technique was used to determine the internal stress present in the coatings, in which an inter-planar spacing "d" serves as an internal strain gauge. The preferential orientation or huge texture of this TiN thin film makes the bi-axial residual stress measurement difficult. Very few $\psi$ orientations will allow to get some diffraction signal. Therefore, the stress evaluation will be only based on a few measurement points and makes the evaluated residual stress suspicious. Evaluation of residual stress in this case was based on the (422) reflection at around $2\theta$ = $125^{\circ}$. Fig.5 shows the intensity distribution of the (422) Miller plane $2\theta$ vs $\psi$. Texture is clearly observed and hence the reduced number of accessible $\psi$ positions for the stress evaluation.  

 \begin{figure}[h]
\includegraphics[width=0.4\textwidth]{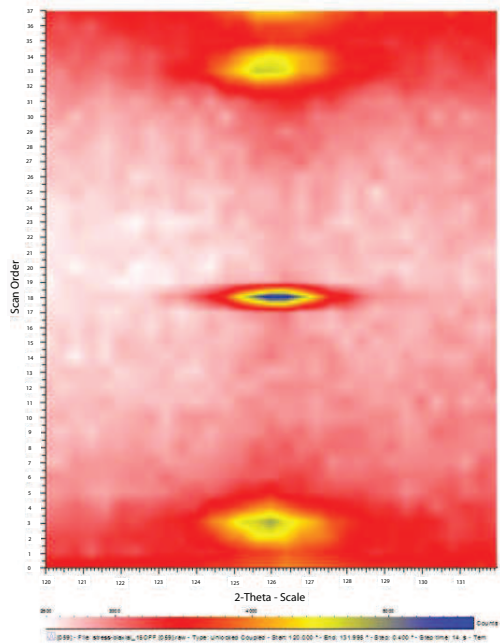}
\caption{\label{Fig5}(Color online) Residual stress mapping Angle ($2\theta$) vs $\psi$}
\end{figure}

\begin{figure}[h]
\includegraphics[width=0.42\textwidth]{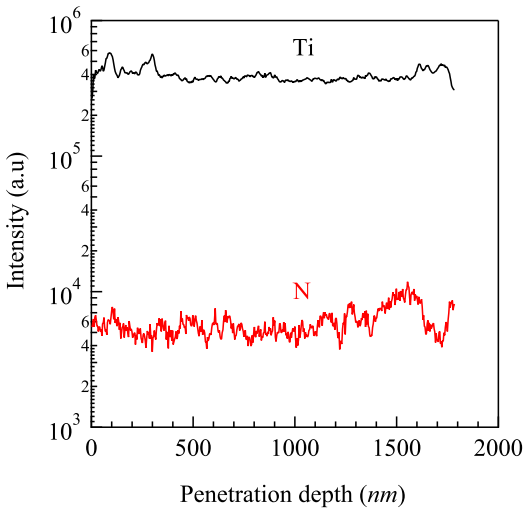}
\caption{\label{Fig6}(Color online) Dynamic SIMS depth profile of TiN film}
\end{figure}

SIMS depth profile of this TiN film is shown in Fig.6. The secondary ion intensities (CsTi$^+$ and CsN$^+$) are markedly different due to dissimilar secondary ion sensitivity factors of the species. The quasi molecular ions of the analyte species with cesium markedly reduce the matrix effect observed in the secondary ion intensity. The elemental analyte intensities and its ratio ($I_{CsTi^+}$ / $I_{CsN^+}$) are constant up to a depth of 1.8~$\mu$m (film thickness) signifying the stoichiometric homogeneity across the depth of the film up to the substrate interface. The intensities of the profiles are matching with TiN standard. However, we cannot calculate the composition quantitatively. Yet the relative intensities of the sample, the standard and their uniformity with respect to depth confirm the sample stoichiometry and its homogeneity. This is significant advantage over many other coating processes where a nitrogen gradient is generated which can lead to sub-stoichiometry and mechanical property gradations.\cite{Ajikumar} The intensity of oxygen and carbon impurities are very low in these thin films. This stems from the localized and high purity selective plasma used in this deposition.

A widely used nanoindentation technique has been used to determine the mechanical properties of e-beam grown TiN thin films. These films have uniform surface morphology with RMS roughness of $\approx$ 5~nm measured by AFM, scanned across an area of 5 x 5~$\mu$m. The low surface roughness is the resultant of low deposition rate which in turn ensured the stoichiometry throughout the film. A typical load - displacement curve obtained for this TiN film is shown in fig.7. The nanoindentation hardness usually depends on the surface roughness, chemical state of surface layer and the size of the indenter. Here, roughness of this film is small enough and it is much lower than the depth penetrated by the indenter. The maximum penetration depth was maintained well within the 10\% of the TiN film thickness. A typical value of indentation hardness and modulus is measured as 19.5~GPa and 214~GPa, respectively. 

\begin{figure}[h]
\includegraphics[width=0.42\textwidth]{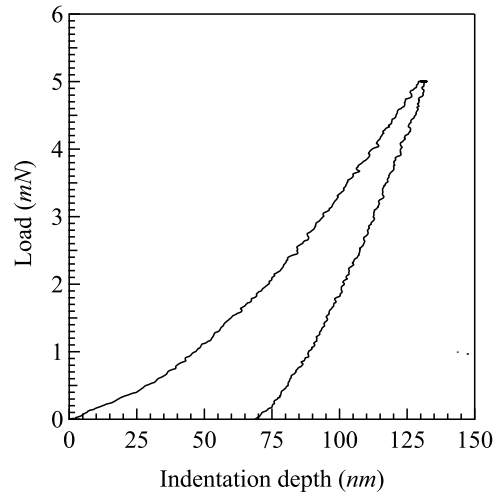}
\caption{\label{Fig7} A typical nanoindentation profile of TiN film}
\end{figure}

\begin{figure}[h]
\includegraphics[width=0.42\textwidth]{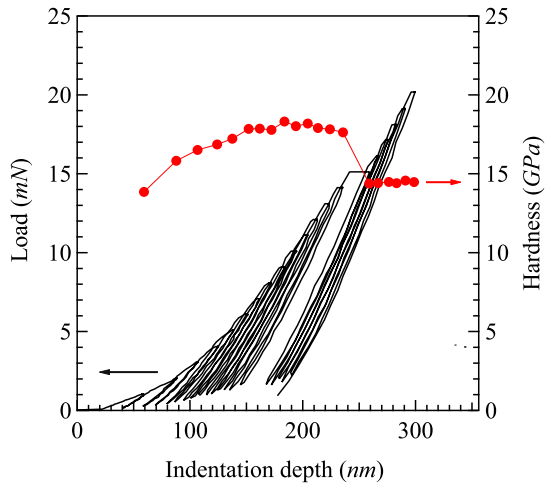}
\caption{\label{Fig8}(Color online) CMC loading profile of TiN film}
\end{figure}

Continuous Multi Cycle (CMC) indentation is a technique that can provide the fatigue information of the thin films. \cite{feby} The load dependent indentation behavior was investigated through CMC testing on the TiN films coated on micro slides (hardness $\approx$ 5~GPa) by increasing the load gradually. The stress field exerted by the indenter is a complex mixture of biaxial stresses includes elements of tension, compression and shear. \cite{Fischer} The reloading paths do not necessarily overlap with the unloading path of the previous loading cycle resulting in hysteresis loops. The reason for this can be ascribed to the chemistry and microstructure of the material led to Bauschinger effect. \cite{Saraswati} Figure 8 shows the continuous loading-unloading indentation profiles with hardness corresponding to penetration depth. Initially, the hardness increases due to the indentation contact in the elastic region \cite{Fischer} and constant hardness range follows. This exhibit the indentation is in the plastic zone (indentation depth range, 150 to 225~nm). The hardness should not vary with depth beyond this range since these films are quite hard. But, at higher indentation depths the residual stress relieving makes the hardness to decrease rapidly. \cite{Bharat, Page} It is understood from the GIXRD profile systematic shift to higher angle supports the existence of compressive residual stress in this film. Usually residual stress is relieved through the formation of crack on the films. But, it is clear from the fig.8 that there is no pop-in or pop-out was observed in the typical indentation profile. This signifies that there is no fracture of films can be expected during this relaxation process. Usually, compressive residual stress makes the films much harder than the stress relieved films. \cite{Lee} 
 
\section{Conclusion}

Stoichiometric single phase fcc TiN films have been successfully synthesized after modification of the e-beam evaporation assembly using an anode plate in the vicinity of the heated titanium ingot to ignite dense plasma, which enhances the reactive deposition. This reaction gets activated further by substrate bias which causes the self-ion bombardment. This plasma enhances ion yield and ion molecule interactions to deposit films at relatively lower temperature. We have obtained a high quality polycrystalline TiN films of grain size $\sim~$150~nm. A $N_2$ jet  focusing induced the preferential orientation in the same direction. Also systematic lattice contraction indicated the presence of compressive residual stress which is beneficial for wear resistance applications. Dynamic SIMS analysis revealed compositional uniformity across the depth and negligible low Z (O, C) impurities. Nano-indentation studies on these films resulted hardness values close to 19.5~GPa. 

\section{Acknowledgement}
F.J would like to acknowledge Mr.A.K.Balamurugan for the discussion regarding the SIMS studies. We would like to thank the UGC-DAE-CSR facility at Kalpakkam for the GIXRD.

\end{document}